\documentclass[conference,letterpaper]{IEEEtran}
\usepackage{fancyhdr,graphicx}
\begin{document}

\title{Reverse method for labeling the information from semi-structured web pages}

\author{\authorblockN{Z. Akbar and L.T. Handoko}
\authorblockA{Group for Theoretical and Computational Physics, Research Center for Physics, Indonesian Institute of Sciences (LIPI)\\
Kompleks Puspiptek Serpong, Tangerang 15310, Indonesia\\
zaenal@teori.fisika.lipi.go.id, handoko@teori.fisika.lipi.go.id}}

\maketitle

\thispagestyle{fancy}
\fancyhead{}
\lhead{}
\lfoot{978--1--4244--2287--6/08/\$25.00~\copyright~2008 IEEE}
\cfoot{}
\rfoot{}
\renewcommand{\headrulewidth}{0pt}
\renewcommand{\footrulewidth}{0pt}

\begin{abstract}
We propose a new technique to infer the structure and extract the tokens of data  from the semi-structured web sources which are generated using a consistent template or layout with some implicit regularities. The attributes are extracted and labeled reversely from the region of interest of targeted contents. This is in contrast with the existing techniques which always generate the trees from the root. We argue and show that our technique is simpler, more accurate and effective especially to detect the changes of the templates of targeted web pages.\\
\end{abstract}


\IEEEpeerreviewmaketitle

\section{Introduction}

\begin{figure*}[t]
 \centering
 \includegraphics[width=16cm]{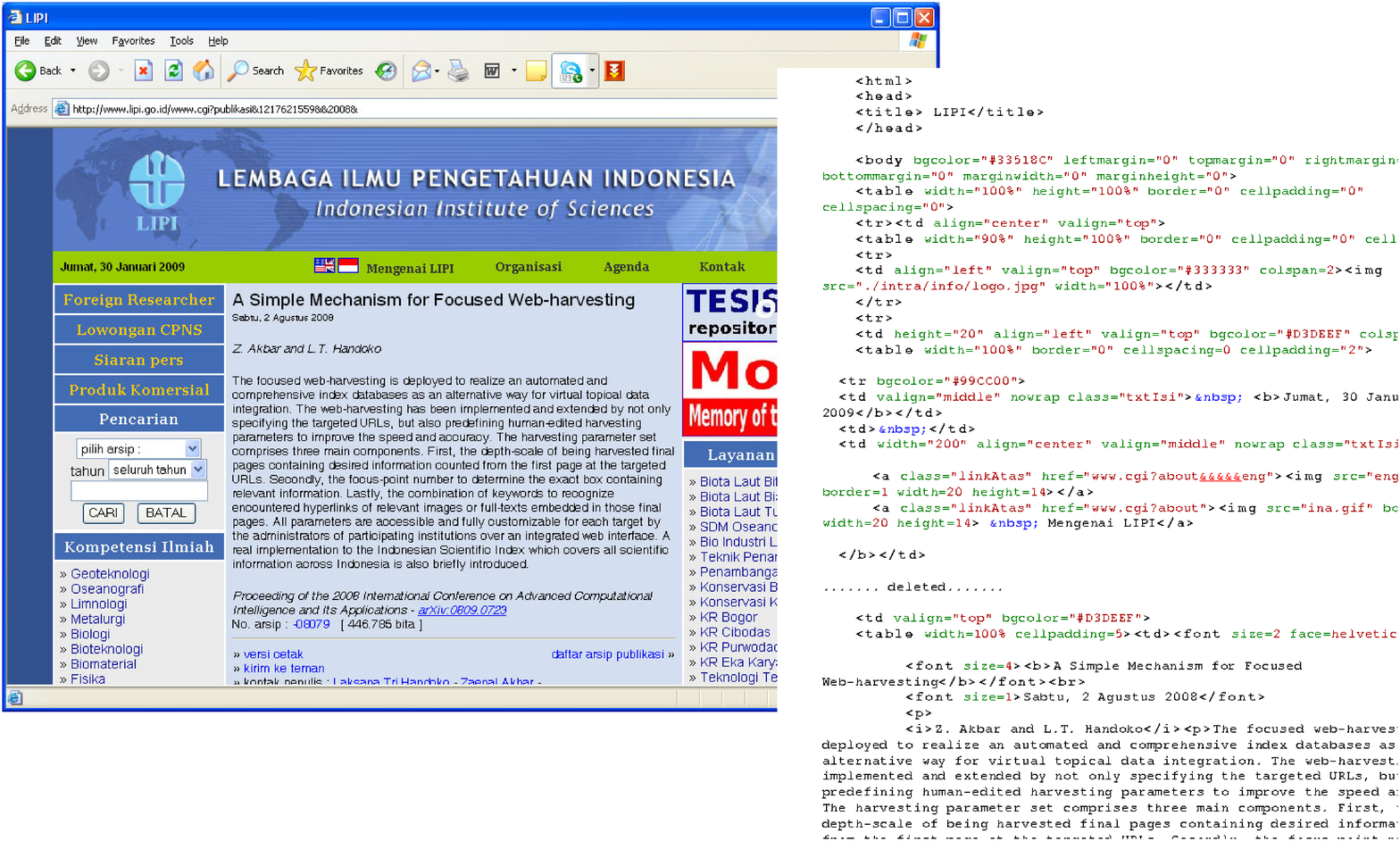}
 \caption{The example of the final RoI page of a publication on the web and its RoI's HTML source.}
 \label{fig:source}
\end{figure*}

During the last decade, most websites are providing the information generated from the structured data in an underlying database through certain predefined templates or layouts. Following the great number of web pages available on the net, these semi-structured web sources contain rich and unlimited valuable data for a variety of purposes. Extracting those data and then rebuilding them into a structured database are a challenge to realize an automatic data mining from web sources. 

Several methods for these purposes have been proposed previously in the literature. Some of them can be classified as the so-called wrappers, for instance \cite{w1,w2,w3,w4,w5}, and have been briefly surveyed in \cite{w6}. The wrapper technique allows an automatic data extraction through predefined wrapper created for each target data source. The wrappers then accepts a quesry against the data source and returns a set of structured results to the calling application. 

On the other hand, there are several automatic methods without a manual initial learning process. For example, some methods are generating a template automatically from first multiple pages before extracting the rest of data based on the template \cite{t1,t2,t3}. A more comprehensive method without requiring multiple pages has also been proposed using a page creation model which captures the main characteristics of semi-structured web pages to derive the set of properties \cite{ti1}. Though, the last method is more intended to extract the lists of data records from a single web page with sibling subtrees.

Recently we have worked on extracting the semi-structured data from  targeted web pages with specific topics, that is the so-called focused web-harvesting \cite{isipaper}. The method is suitable for in particular the 'indirect data integration' which is not tolerant to any error. The architecture is inspired and the combination of focused web-crawling and regular web-harvesting. The focused web-crawling does not indiscriminately crawl the web pages like general purpose search engines, but attempts to download pages that are similar to each other \cite{chakrabarti}. According to the data integration purposes and its requirement of high accuracy, the focused web-harvesting technique adopts human intervention in the initial setup by providing the targeted URL of list of data, for instance the publication list, and defining the template of the final page contains the relevant information by labeling the attributes. In the case of detail information of publication page as shown in Fig. \ref{fig:source}, the relevant information and labeling are  ranging from title, authors, abstract to the fulltext if available. The method has been applied to the Indonesian Scientific Index (ISI) which integrates the scientific data from scientific and academic institutions across Indonesia through their websites \cite{isi}. The system is also open for public under GNU License at SourceForge.net \cite{openisi}. Obviously, in spite of its high accuracy and unnecessary machine-learning like system, the method is suffered from tedious labor time at its initial setup to determine the tokens, to label the attributes, and is lacking of the ability to detect effectively later changes of the targeted page templates. Because the labels and tokens are represented as DOM trees which are sensitive to later changes of targeted web templates \cite{label,dom}. 

In this paper, in order to overcome the above-mentioned problems we present a new method to extract the tokens reversely from the region-of-interest (RoI) at the final web pages, and further label each attribute as normal procedure. This technique is in contrast with the existing mechanism which always starts from the root of web source and also the root of RoI. We argue that it is more efficient and accurate to extract the tokens, and on the other hand to detect the later template changes withour any ambiguities. We should also remark that the reverse method is applicable for any existing methods for data extraction, especially the ones which require initial setup by human intervention to define and label the attributes. This is actually similar to the previous method \cite{chang}, but instead of using the PAT tree \cite{pat} we use the DOM tree like mechanism \cite{dom}. Moreover we improve its accuracy by taking into account the lower part of tree and not only the trees from the root.

The paper is organized as follows. First, after this brief introduction we describe our approach in Sec. \ref{sec:tree}. In the subsequent section, Sec. \ref{sec:implementation}, we present the implementation and the web interface to define the initial setup for each targeted web. Finally we summarize the paper and provide some future issues and further development.

\section{The reverse mechanism}
\label{sec:tree}

\begin{figure*}[t]
 \centering
 \includegraphics[width=16cm]{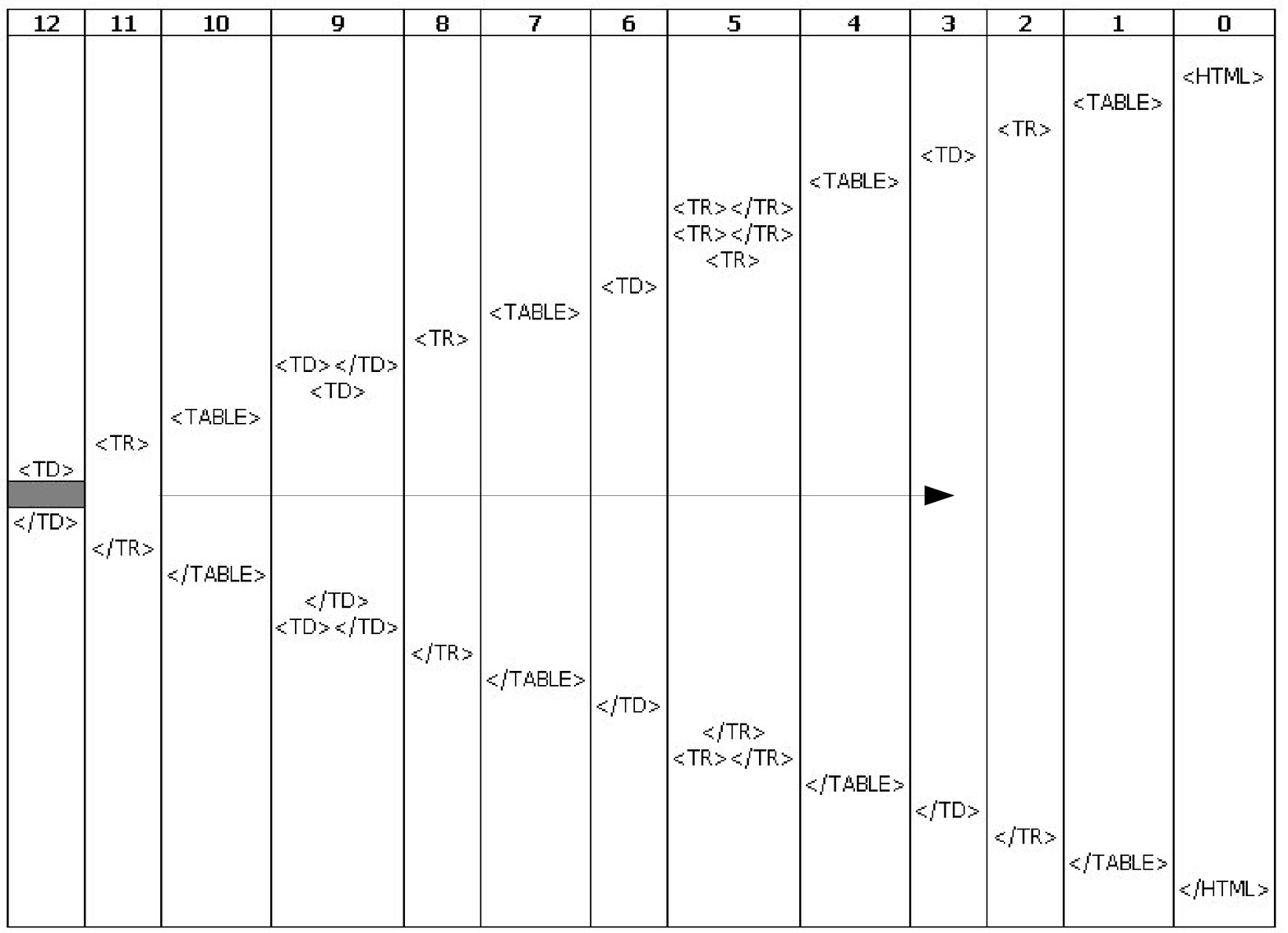}
 \caption{The tree for the example of RoI given in Fig. \ref{fig:source}. The dashed box denotes the RoI.}
 \label{fig:tree}
\end{figure*}

No matter the method used to extract and to label the tokens from a web template or layout, correct initial setup is crucial for further data extraction. As mentioned before, this point plays an important role for indirect data integration which has no tolerance to any errors. This makes some methods based on the machine learning system are useless. 

From now, please note that we are not going to deal with the algorithms to mine the labeled data since the tasks after labeling can be done further using any existing methods, nor to find the relevant pages of data list which has been discussed in our previous work \cite{isipaper} and many previous works elsewhere. The reverse mechanism can be outlined recursively as follows :
\begin{enumerate}
\item Determine the URL of the final web page with desired information like Fig. \ref{fig:source}.
\item Provide the whole sentences of the RoI by copying and pasting the 'displayed desired text'.
\item Provide the whole sentence(s) of each sub-RoI and assign the attributes for each of them.
\item Crawl the source.
\item Parse and clean the text-format HTML tags like \verb|<b>|, \verb|<i>| etc.
\item Take the upper part of source from the top till the last one before the first sentence of RoI. Parse and clean all texts inside except the layout-format HTML tags, like \verb|<tr>| etc, Do the same thing for the lower part that is from the end of last sentence of RoI till the bottom.
\item Calculate the number of 'open-tag' ($n_\mathrm{ot}$) and 'closed-tag' ($n_\mathrm{ct}$) from the deepest part in term of desired content, that is the nearest tags from the RoI. 
\end{enumerate}
We should stress here that there is no need for the administrators to provide the web page sources at all. Open-tag here means the tags which have no pair in upper or lower part, while the closed-tag denotes the pairing tags within upper or lower part. Of course, our interest is only in the open-tag which should describe the whole structure of web template.

Following the above procedures, we can obtain a kind of DOM tree as shown in Fig. \ref{fig:tree}. We can further calculate the number of trees according to the number of open-tags. Please remind that the calculation is done horizontally, from left to right shown by the arrow in the figure. The number of trees in upper and lower parts are determined by, 
\begin{equation}
 \Sigma \equiv n_\mathrm{ot} - n_\mathrm{ct} \, .
\end{equation}

Concerning all possibilities on the number of trees in upper and lower parts, therefore we can generally categorize the web structures through the discrepancies between both numbers as,
\begin{eqnarray}
 \Delta & & \equiv \Sigma_\mathrm{upper} - \Sigma_\mathrm{lower} 
\nonumber \\
 & & \left\{ 
\begin{array}{lcl}
 = 0 	& : & \mathrm{fully \; symmetry} \\
 < 0 	& : & \mathrm{lower \; asymmetry} \\
 > 0 	& : & \mathrm{upper \; asymmetry} \\
\end{array}
\right.
\end{eqnarray}
Fig. \ref{fig:tree} provides an example of tree in the case of Fig. \ref{fig:source} which is accidently asymmetry. That means the number of trees in the upper and lower parts are not the same, $\Sigma_\mathrm{upper} \neq \Sigma_\mathrm{lower}$. Again, we can use one of the existing methods to calculate the number of trees like  the PAT tree algorithm \cite{pat} and so forth. 

Through the discussion above, it is clear that the present method has several advantages :
\begin{itemize}
 \item We can separate independently the structure and the rules to obtain the RoI and the structure inside. 
\item We can find out the template changes and its relevance with the desired RoI, since we can compare and see the pairing tokens between the upper and lower parts.
\end{itemize}
We discuss these points in more detail through the real implementation at ISI in the subsequent section.

\section{The implementation}
\label{sec:implementation}

\begin{figure*}[t]
 \centering
 \includegraphics[width=16cm]{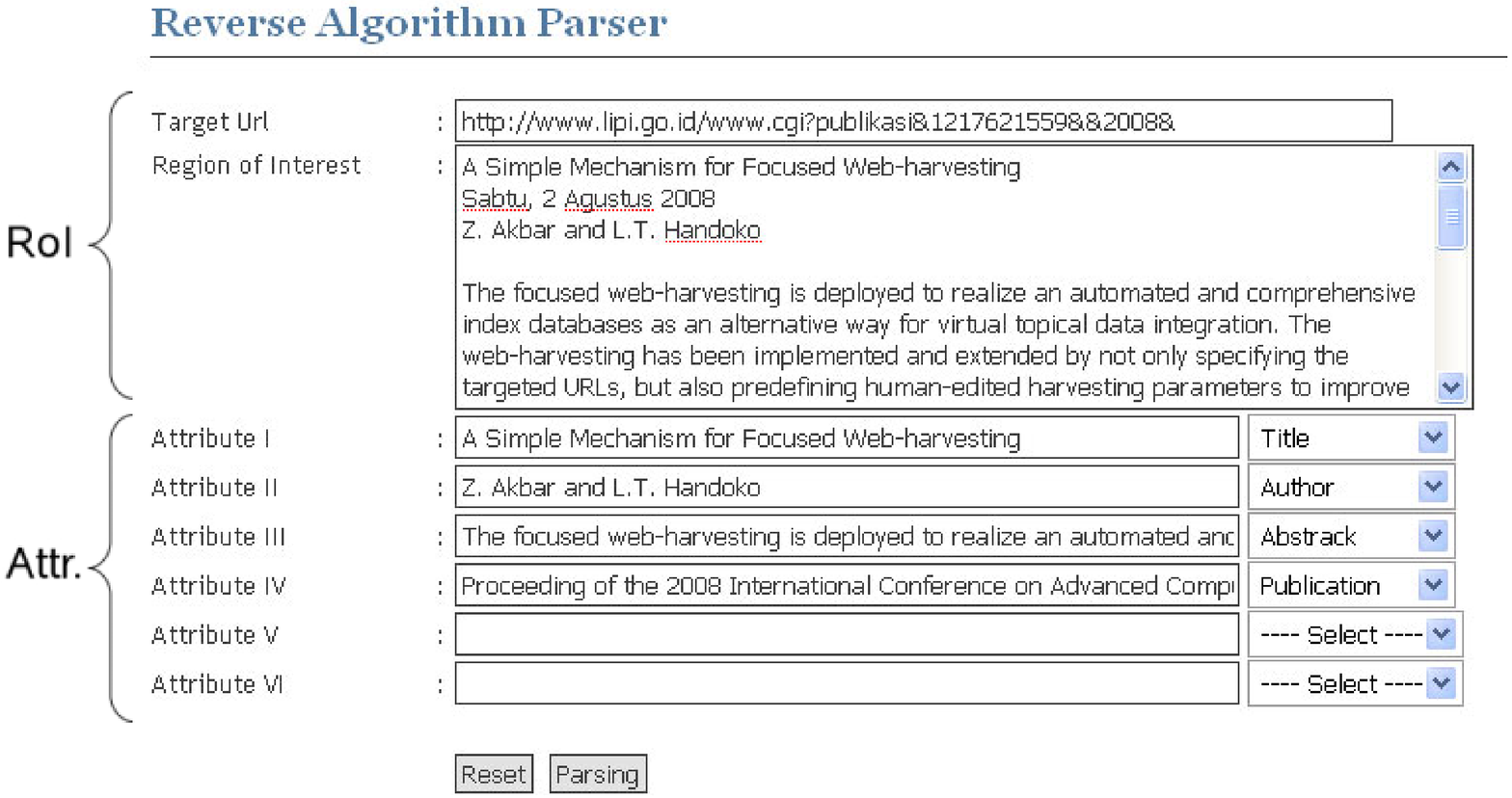}
 \caption{The web interface to define the RoI and to label the attributes.}
 \label{fig:gui}
\end{figure*}

Our approach to web page information extraction has been experimentally implemented into the system of ISI. Following the wrapper induction programs which are usually supplemented by a user friendly GUI, we have also developed a web based interface to perform the initial setup. The example used at ISI is given in Fig. \ref{fig:gui} which shows the web interface to define the content of RoI from a choosen final page and the labels for each attribute.

ISI can now efficiently detect the changes of targeted web templates by comparing the old and new numbers of $\Delta$. It is done by executing the check procedure each time prior to the new crawling works into the same targeted web pages. Since the parameter as in Fig. \ref{fig:gui} has been stored in the system, it can be used not only for the initial setup but also for rechecking the templates in a regular basis.

The discrepancies between the old and new numbers of $\Delta$ is usefull to detect easily the template changes time by time, and at the same time determine where the changes happened. It can be summarized as below,
\begin{enumerate}
 \item No change at all :\\
	$\Delta^\mathrm{new} = \Delta^\mathrm{old}$; 
	$\Sigma_\mathrm{upper}^\mathrm{new} = \Sigma_\mathrm{upper}^\mathrm{old}$; 
	$\Sigma_\mathrm{lower}^\mathrm{new} = \Sigma_\mathrm{lower}^\mathrm{old}$
\item Simultaneous changes with same size in both upper and lower trees : \\
	$\Delta^\mathrm{new} = \Delta^\mathrm{old}$; 
	$\Sigma_\mathrm{upper}^\mathrm{new} \neq \Sigma_\mathrm{upper}^\mathrm{old}$; 
	$\Sigma_\mathrm{lower}^\mathrm{new} \neq \Sigma_\mathrm{lower}^\mathrm{old}$ 
\item Only one tree has changed :\\
	$\Delta^\mathrm{new} \neq \Delta^\mathrm{old}$; 
	$\Sigma_\mathrm{upper}^\mathrm{new} \neq \Sigma_\mathrm{upper}^\mathrm{old}$; 
	$\Sigma_\mathrm{lower}^\mathrm{new} = \Sigma_\mathrm{lower}^\mathrm{old}$ \\
	or :  \\
	$\Delta^\mathrm{new} \neq \Delta^\mathrm{old}$; 
	$\Sigma_\mathrm{upper}^\mathrm{new} = \Sigma_\mathrm{upper}^\mathrm{old}$; 
	$\Sigma_\mathrm{lower}^\mathrm{new} \neq \Sigma_\mathrm{lower}^\mathrm{old}$ \\
\item Both trees have changed differently :\\
	$\Delta^\mathrm{new} \neq \Delta^\mathrm{old}$; 
	$\Sigma_\mathrm{upper}^\mathrm{new} \neq \Sigma_\mathrm{upper}^\mathrm{old}$; 
	$\Sigma_\mathrm{lower}^\mathrm{new} \neq \Sigma_\mathrm{lower}^\mathrm{old}$ \\
\end{enumerate}
Apparently, in the case 1 no need to alter the saved intial parameter. In contrast, from the case 2, 3 and 4 we can deduce that the templates have been changed, either in the upper, lower or both trees. The important point is, once the template changes have been detected, the system automatically replace the old version of template with the new one.

Furthermore, the reverse method can be used to rebuild the content of RoI in terms of labels which enable us to restructure the database for further purposes. The procedure is completely the same as extracting the RoI. Tab. \ref{tab:attr} shows the delimiters extracted from the RoI in Fig. \ref{fig:source} using the same interface as Fig. \ref{fig:gui}.

\begin{table}[b]
\centering
\begin{tabular}{|c|c|c|}
	\hline
	Start 	& Attributes 	& End 	\\
	\hline
		& Title		& \verb|<br>|	\\
	\verb|<p>| 	& Author	&  \\
	\verb|<p>|	& Abstrack	&	\\
	\verb|<p>|	& Publication	& \verb|<br>| \\
	\hline
\end{tabular}
 \caption{The delimiters extracted from the example in Fig. \ref{fig:source} using the reverse mechanism defined in Fig. \ref{fig:gui}.}
 \label{tab:attr}
\end{table}

\section{Summary}

We have discussed a simple method based on the reverse algorithm and DOM tree to extract the RoI and label the relevant attributes in the initial setup. The resulted patterns can be used further to automatically extract the data from crawled targeted web pages. We argue that the method and its web interface reduce the administrator works significantly, while on the other hand improve the accuracy and speed of finding the tokens and labeling the attributes. We have found that this method is very effective to detect the template changes, for instance newly inserted advetorial in the middle of upper or lower tree which often occurs in any websites and leads to difficulties in existing methods.

The experimental works on applying the method to the available huge number of data stored at ISI is still under progress. The expected results and its effectiveness to detect the altered web pages will be analysed and published in a more complete and detail paper elsewhere. However, according to our trial experiments using 10000 data from several web sources, the method performs very well. It succeeded in detecting any template changes and improved the speed of whole processes up to 20\%.

Finally, we should remark that in principle the method is applicable for the web sources in a form of list of data. The work on this matter is also in progress.

\section*{Acknowledgment}

The work is partially supported by the Riset Kompetitif LIPI in fiscal year 2009.

\end{document}